# An Energy-Autonomous and Battery-Free Resistive Sensor using a Time-Domain to Digital Conversion with Bluetooth Low Energy connectivity

Mario Costanza*, Antonino Pagano‡§, Samuel Margueron *, Ilenia Tinnirello ‡§¶, Roberto La Rosa†

* FEMTO-ST Institute, University of Franche-Comte, ENSMM, CNRS UMR6174, Besanc¸on, 25030, France´ Email: mario.costanza@femto-st.com

† STMicroelectronics, Stradale Primosole 50, Catania 95121, Italy

‡ Department of Engineering, University of Palermo, Viale delle Scienze, Palermo, Italy

§ CNIT - Consorzio Nazionale Interuniversitario per le Telecomunicazioni, Parma, Italy

¶ Department of Electrical, Electronics and Computer Science Engineering, University of Catania, Italy

*Abstract*—This paper introduces an innovative EnergyAutonomous Wireless Sensing Node (EAWSN) that addresses power constraints by harnessing ambient light for energy. It combines this energy harvesting capability with the Time Domain to Digital Conversion (TDDC) technique for efficient and accurate measurements of resistive sensors. Bluetooth Low Energy (BLE) communication ensures data can be transmitted wirelessly to a base station, providing a promising solution for various applications, particularly in environments with limited access to wired power sources, enabling long-term, maintenance-free operation by eliminating batteries. Experimental results showed a linear relationship between the test resistance $R_m$ and the measured number of clock pulses $N_m$ within the sensor's operating range.

## I. INTRODUCTION

Distribute sensing of a defined environment is essential in controlling complex, large-scale systems with the objective of higher efficiency and reliability. In such systems, a wired approach for power delivery and inter-device communication between several nodes is often unfeasible or undesirable, for example, because of an already existing network or the added costs and complexity from the cables required. From this perspective, state-of-the-art networks typically employ multiple battery-powered sensing nodes communicating through a practical wireless protocol (Bluetooth, Zigbee, LoRa, and the like). However, a battery-powered Wireless Sensor Node (WSN) has the drawback that the batteries need periodical maintenance, which is time-consuming and waste-producing [1]. In addition, the total cost of the batteries can easily overshadow the cost of the single device over time, and to solve this issue, the use of

This work was partially supported by the following projects: - The ANR within the framework of the EUR EIPHI project (ANR-17-EURE-0002).
- The EU under the NRRP of NextGenerationEU, partnership on "Telecommunications of the Future" (PE00000001 - program "RESTART"). - The EU Horizon Europe research and innovation programme under the Marie Skłodowska-Curie grant agreement No. 101086359.

various strategies can extend the lifespan of batteries. These strategies primarily involve reducing the power consumption of the devices or reintegrating energy through methods such as Energy Harvesting (EH) or Wireless Power Transfer (WPT). Principal strategies for reducing the power consumption are, a careful optimization of the programming of the devices, improvements in the architecture, and usage of lower-power sensors. In litterature, a battery-less architecture has already discussed how it is possible with an EH approach within the few-μW power range in the works of *La Rosa et al.* [2], [3]. Further, the specific choice of the sensors and relative readout also cares attention for the optimal implementation of a WSN [4], [5]. In this context, this paper shows a technique for Time-based measurements through Time Domain to Digital Conversion (TDDC) techniques [6], [7]. These techniques encode a measured quantity as a time and can usually be energy-efficient and accurate. TDDC techniques have been used in a variety of applications, like LiDARs [8], [9], temperature sensing [10], [11], time-domain ADCs [12], ADPLL (all digital phase-locked loop) [13], resistance measurements [14], capacitive measurements [15]. Resistancebased sensors provide a cheap and accurate means to measure a variety of physical phenomena [16]–[18], and their combination with TDDC techniques can have a significant impact on the overall energy efficiency of wireless nodes. This work explores an innovative Battery-Free Wireless Resistive Sensor Node (EAWRSN) through a TDDC conversion. The energy-autonomous nature of the system is achieved through a small solar cell and the usage of a small capacitor to store the harvested energy, a Bluetooth Low Energy (BLE) SoC for wireless communication, and an ultra-low-power microcontroller for power management and carries out the TDDC measurements. Researchers have proposed a similar architecture, demonstrating sensor readings via a time domain readout of, for example, vibrations [19] or ambient light monitoring [4]. Although the hardware used in this article is the same, the concept of resistive sensor reading with TDDC is different, and for this purpose, an innovative firmware was developed. For instance, in the paper [19], the sensor exploited the proportionality between the time elapsed between two successive beacons (advertising time) and the acceleration imparted by the piezoelectric harvester. Additionally, in their work, the advertising time was measured remotely from a Base Station (BS) during the reception of beacons, and then the acceleration was indirectly calculated. In contrast, our solution proposes the resistive sensor readout technique based on TDDC, in which the time measurement is implemented directly in the microcontroller. This work is organized as follows. Section II describes the system in terms of behavior, key parameters, and circuit implementation. Section III deals with the TDDC resistive sensing technique. Section IV describes the system setup and shows the experimental results. Section V concludes the paper.

## II. System Description

Fig. 1 shows the architecture of the EAWSN. The Energy-Autonomous feature of the system is achieved through an Energy Harvesting Unit (EHU) that harvests the energy stored in a small capacitor $C_{stor}$ and monitored through the voltage $V_{stor}$. The harvested energy $P_{geh}$ is managed by the System and Energy Control Unit (SECU), which directs it to the various components of the system, gathers the measurements through the sensors and delivers the information to a remote BS through the radio. The BS has the same connectivity as the EAWSN, is powered via a stable power source or a battery and can perform more energy-intensive operations.

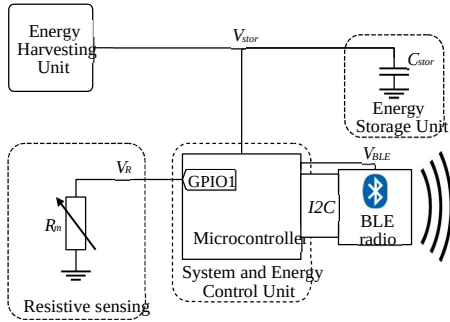

Fig. 1. Block diagram of the resistance measurement system.

This specific architecture allows the system to operate under very low (few $\mu W$) harvested power. As shown in Fig. 2, the system operation is subdivided into low-power harvesting phases and high-power active phases. During the harvesting phase, the system is configured to minimize the quiescent power $P_q$, allowing $C_{stor}$ to charge up to a defined threshold $V_H$, triggering an active phase. An active phase represents any energy-intensive operation (sending a packet, performing a measurement, et cetera) that causes an imbalance of power that is reflected in the discharge of the capacitor $C_{stor}$, corresponding to a defined voltage drop.

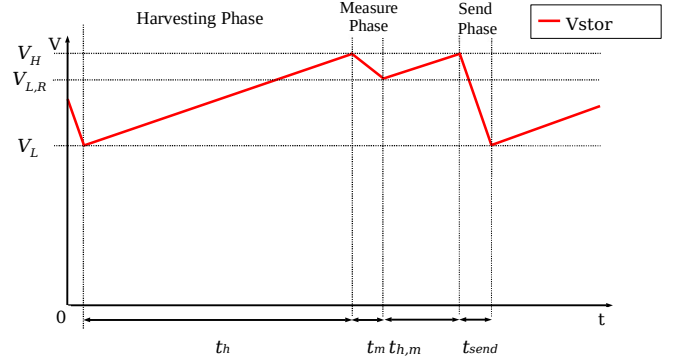

Fig. 2. Typical work cycle of the EAWRSN.

The time duration of the harvesting phase $t_h$ is measured by the SECU through an internal timer and an internal reference $f_{clk}$, as described in equation (1). Physically, $t_h$ is defined by the value of $C_{stor}$, the energy consumed during the previous active phase $E_{act}$ and the difference between $P_{geh}$ and $P_q$, as defined in equation (2).

$$t_h = \frac{N_h}{f_{clk}} \quad (1)$$

$$t_h = \frac{E_{act}}{P_{geh} - P_q} \quad (2)$$

The equations (1) and (2) allow to estimate the $P_{geh}$ through a TDDC technique using $N_h$, as demonstrated in equation (3). This technique enables a secondary usage of the harvester as a sensor, as already shown, for light intensity [2] and vibration intensity [20].

$$P_{geh} = \frac{E_{act}}{t_h} + P_q = \frac{E_{act} \cdot f_{clk}}{N_h} + P_q \quad (3)$$

The energy requirements of both the active and harvesting phases are the starting point to determine the other system parameters. The highest possible consumed power during an active phase $E_{act\ max}$ specifies the minimum value of $C_{stor}$ through the minimum operating range $V_{L\ min}$ of the devices used as defined in (4).

$$C_{stor} \geq \frac{2 \cdot E_{act\_max}}{V_H^2 - V_{L\_min}^2} \quad (4)$$

The value of $P_q$ during the harvesting phase determines which harvesters may be used, as a reasonable minimum power $P_{geh\ min}$ has to be provided during the defined standard operation, as defined in the condition in (5).

$$P_{geh\ min} \gg P_q \quad (5)$$

The EAWSN embeds uniquely commercial off-the-shelf (COTS) devices. The SECU uses a COTS Cortex-M0+ microcontroller from STMicroelectronics (STM32L031F4) [21]. The SECU delivers the power from $C_{stor}$ to the connected devices via the on-board General Purpose Input/Output (GPIO) and performs time measurements via the Low-Power Timer (LPTIM). Finally, the harvested energy is observed through the $V_{stor}$ using the internal Programmable Voltage Detector (PVD), which is a software-programmable multilevel comparator, with granularity 200mV over the range 3.2V - 2.0V. During the harvesting phase, the microcontroller works in stop mode, and the total quiescent current is 1μA, corresponding to a quiescent power $P_q$ of ~2.5μW.

Given the low value of the power $P_q$, a small (1.5cm×1.5cm) Panasonic AM-1606C amorphous silicon photovoltaic cell was enough, as it provides ≈ 10μW for a light intensity of 200lux and temperature 25°C, at the defined optimal operating point [22]. The radio used Bluetooth-Low-Energy (BLE) to both satisfy the low-energy system requirement and have high compatibility with commercially available devices like computers, smartphones, and tablets. Specifically, the system embeds the BlueNRG-2 from STMicroelectronics configured to send beacons at a transmission power of +8dBm. The capacitor $C_{stor}$ was chosen considering the most energy-intensive action of the EAWSN, which was sending the beacons. Under that condition, $C_{stor}$ connects two standard capacitors with a total nominal value of 440μF. This value was experimentally determined considering a target $V_L$ of ~2.0V to maintain some margin on the defined minimum operating voltage of 1.8V.

### III. RESISTIVE SENSING TECHNIQUE

As displayed in Fig. 3, the technique implemented for measuring the resistor $R_m$ makes optimal use of the battery-free nature of the EAWSN to encode the value of $R_m$ in the time $t_m$ needed to discharge the storage capacitor $C_{stor}$ from the threshold $V_H$ to a defined value $V_{L,R}$ using $R_m$ itself as load. The resulting architecture has the immediate advantage of simplicity since it only requires one electrical connection to the measurand. Furthermore, as $V_H$ and $V_{L,R}$ are set, the energy consumed during the measurement $E_m$ is a constant value defined as in (6). As $C_{stor}$ is set through (4), this highlights that to reduce $E_m$, $V_{L,R}$ should be as close as possible to $V_H$.

$$E_m = \frac{1}{2} \cdot C_{stor} \cdot \left(V_H^2 - V_{L,R}^2\right) \quad (6)$$

Fig. 1 shows that to perform the measure, the microcontroller starts the LPTIM, sets the digital output GPIO1 to 1, and goes into the low-power stop mode. When the measurement ends through the wake-up event from the PVD, when $V_{stor} = V_{L,R}$, the microcontroller stores the counted value $N_m = t_m \cdot f_{clk}$ to send it at the BS during the next send phase. Finally, from the knowledge of the operating parameters of the EAWSN, the BS computes the value of $R_m$.

When the measurement is started, $V_R$ is connected to $V_{stor}$ and the power provided from $C_{stor}$ to be the one described in (7).

$$P_{Cstor} = P_R + P_q - P_{geh} \quad (7)$$

Through the condition expressed in (5) and considering that, during the measure, $P_R = V_{stor}^2/R_m$, equation (7) is rewritten as (8).

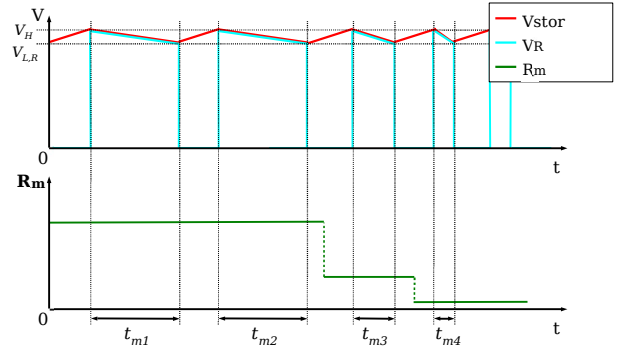

Fig. 3. $V_{stor}$ and $V_R$ evolution during the resistor measurement phase.

$$P_{Cstor} = \frac{V_{stor}^2}{R_m} + P_q - P_{geh} \quad (8)$$

By restricting to the validity of the condition 9, equation (8) defines the exponential discharge of a capacitor, and $t_m$ is found as in equation (10).

$$\frac{V_{L,R}^2}{R_m} \gg |P_{geh} - P_q| \quad (9)$$

$$t_m \approx R_m \cdot C_{stor} \cdot \ln\left(\frac{V_H}{V_{L,R}}\right) \quad (10)$$

Finally, as $t_m$ is read digitally through the number of clock pulses $N_m$ returned by the timer LPTIM using the reference $f_{clk}$ as in (11), the value of $R_m$ is defined in function of $N_m$ via the equation (12).

$$t_m = \frac{N_m}{f_{clk}} \quad (11)$$

$$R_m = \frac{N_m}{f_{clk} \cdot C_{stor} \cdot \ln\left(\frac{V_H}{V_{L,R}}\right)} \quad (12)$$

The range of validity for the method proposed is described through a maximum value $R_{max}$ and a minimum $R_{min}$. A mathematical description of $R_{max}$ is found through (9) as in (13), while $R_{min}$ defined via (12) as (14).

$$R_{max} \ll \frac{V_{L,R}^2}{P_{geh}} \quad (13)$$

$$R_{min} > \frac{2}{f_{clk} \cdot C_{stor} \cdot \ln\left(\frac{V_H}{V_{L,R}}\right)} \quad (14)$$

## IV. Experimental Results

The EAWSN has been tested to perform the resistive measurements as explained in section III. The device was connected in the measurement setup shown in Fig. 4 to validate the TDDC resistance measurement technique. The value of $P_{geh}$ was controlled through a small LED powered through a power supply to provide the same power in each measurement session, and the TDDC technique was tested using various resistance values provided via the PRS-330 Precision Programmable Resistance Box [23].

The measured data were recorded simultaneously through Bluetooth, using a remote computer connected to the BS, and by probing the signals $V_R$ and $V_{stor}$ through an oscilloscope.

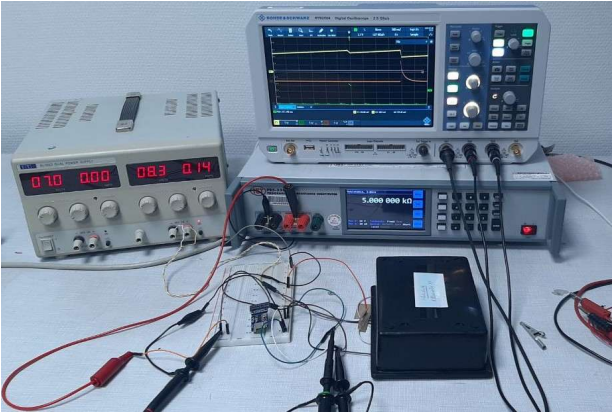

Fig. 4. Experimental setup.

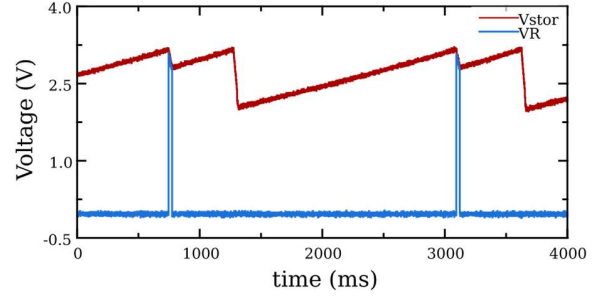

Fig. 5. Evolution of the voltage $V_{stor}$ and $V_R$ vs. time for a test resistance of 1kΩ, resulting in a reading of $t_m$ = 31ms.

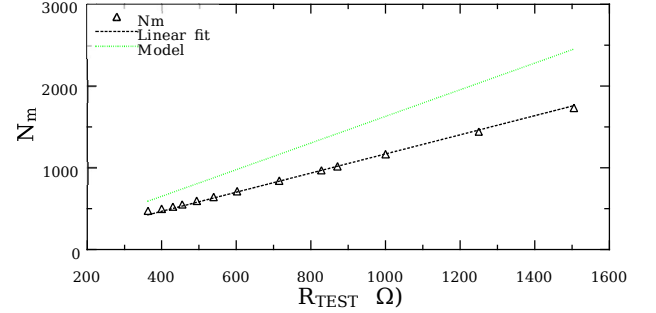

Fig. 6. Measured number of clock pulses and uncalibrated model vs. applied test resistance values.

The EAWSN was configured with the following parameters: (i) $V_H$ = 3.15V; (ii) $V_{L,R}$ = 2.85V; (iii) $f_{clk}$ = 37kHz. Fig. 5 displays a typical measurement cycle captured by the oscilloscope for a test resistor of 1kΩ and an illuminance of 1900lux, which was equivalent to a harvested power $P_{geh}$ of ≈1mW. The data in Fig. 6 show the relation of the measured $N_m$ against the test resistance $R_{TEST}$. As described in section III, the data is described by a line. However, the slope of the curve obtained through (12) (1.629 Ω$^{-1}$) is different than what was obtained by interpolating the measurements assuming a response of the same type (1.169Ω$^{-1}$). Lastly, the data in Fig. 7 plots the value $R_m$ estimated by the BS using equation (12) and the interpolated response using the data from Fig. 6 vs. the test resistance $R_{TEST}$, the two estimations show an $R^2$ factor of ~0.4518 and ~0.9968 respectively. The relative error values, available in Fig. 8, are within ±32 % for the modelled response and ±12 % for the linear fit. The large difference between the values obtained through (12) and the measures is to be attributed from a combination of the 20% uncertainty of $C_{stor}$, a nonzero output resistance of the GPIO and uncertainty on the value $V_H$ and $V_{L,R}$. A specific calibration process on the EAWSN itself could be implemented to systematically correct this error.

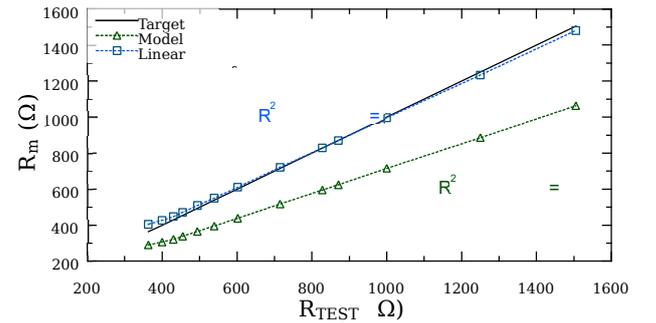

Fig. 7. Estimated resistance vs. applied test resistance values.

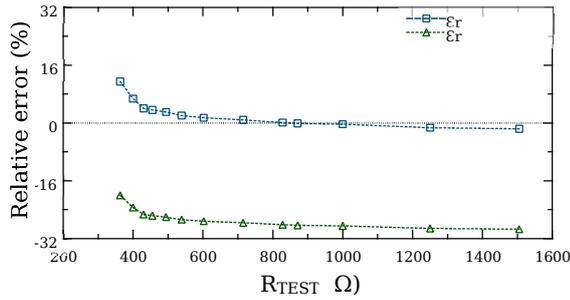

Fig. 8. Relative error of estimated resistance vs. applied test resistance values.

V. CONCLUSIONS

This work presents the development and evaluation of energy-autonomous wireless sensor nodes equipped with Time Domain to Digital Conversion (TDDC) capabilities. The system is autonomous through an energy harvesting approach by converting the ambient light into electrical energy, allowing for self-sustaining and maintenance-free operation, eliminating the need for external power sources like batteries. The TDDC approach encodes the resistance as a function of the discharge time of the storage, enabling a simple and energy-efficient architecture. The converted digital value is sent through Bluetooth Low Energy (BLE) to a remote Base Station (BS), making the resulting device ideal for deploying a true set-and-forget and easy-to-upgrade sensing network that is compatible with a variety of environments. The characterization of the TDDC results aligns with the theory of the device. However, the uncertainty on the storage capacitor and other secondary effects limit the accuracy of the results, indicating the necessity of a calibration process to improve the measurement system.